\documentclass{Interspeech}

\interspeechcameraready

\title{Generating Diverse Audio-Visual 360º Soundscapes for Sound Event Localization and Detection}

\author[affiliation={1}]{Adrian S.}{Roman}
\author[affiliation={1}]{Aiden}{Chang}
\author[affiliation={2,3}]{Gerardo}{Meza}
\author[affiliation={4}]{Iran R.}{Roman}

\affiliation{}{University of Southern California}{United States of America}
\affiliation{}{Universidad Nacional Autónoma de México }{Mexico}
\affiliation{}{Universidad Autónoma de Tamaulipas}{Mexico}
\affiliation{}{Queen Mary University of London}{United Kingdom}
\email{contact: i.roman@qmul.ac.uk}

\usepackage{comment}

\usepackage{amssymb}

\usepackage[table]{xcolor}
\definecolor{lightred}{RGB}{255,204,203}
\definecolor{mediumred}{RGB}{255,153,153}
\definecolor{darkred}{RGB}{255,102,102}
\definecolor{lightblue}{RGB}{204,229,255}
\definecolor{mediumblue}{RGB}{153,204,255}
\definecolor{darkblue}{RGB}{51,153,255}

\usepackage{tikz}

\begin{document}

\maketitle

\begin{abstract}

\noindent We present SELDVisualSynth, a tool for generating synthetic videos for audio-visual sound event localization and detection (SELD). Our approach incorporates real-world background images to improve realism in synthetic audio-visual SELD data while also ensuring audio-visual spatial alignment. The tool creates 360$^\circ$ synthetic videos where objects move matching synthetic SELD audio data and its annotations. 
Experimental results demonstrate that a model trained with this data attains performance gains across multiple metrics, achieving superior localization recall (56.4 LR) and competitive localization error (21.9$^{\circ}$ LE). We open-source our data generation tool for maximal use by members of the SELD research community. 

\end{abstract}

\section{Introduction}

Sound event localization and detection (SELD) combines spatial sound source localization with event classification using ambisonic or multichannel audio \cite{adavanne2018sound}. Recent advances address real-world deployment \cite{politis2022starss22}, overlapping sources of the same category \cite{shimada2022multi}, and distance estimation \cite{kushwaha2023sound,liang2023reconstructing}, with applications ranging from assistive technologies \cite{pandya2021ambient} to autonomous navigation \cite{chen2022soundspaces}. The audio-visual extension integrates visual object detection \cite{chen2021localizing}, enabling solutions to track occluded sounding objects in 360° video \cite{shimada2023starss23}, sound origin differentiation \cite{mahmud2024t}, and visual sound tracking even, when using basic audio formats like stereo or mono \cite{mo2022localizing}.

Most SELD systems rely on data-driven approaches \cite{grumiaux2022survey}, but real-world dataset collection remains a challenge \cite{politis2022starss22,shimada2023starss23,evers2020locata, fuentes2021soundata}. Synthetic audio datasets have emerged as effective training tools \cite{roman2024spatial}, demonstrating significant performance benefits \cite{politis2022starss22,roman2024spatial}. For audio-visual SELD, current synthetic methods spatialize stock media to match audio events, but the videos use empty black backgrounds \cite{roman2024enhanced}.

We present an enhanced synthetic data pipeline incorporating naturalistic CC 4.0-licensed background images and a user-defined set of image and video events. Our method improves visual realism while maintaining precise audio-visual alignment, demonstrating measurable performance gains by a SELD model trained with this data, across multiple metrics. The dataset and tools are publicly available to support multi-modal SELD research\footnote{\url{github.com/adrianSRoman/SELDVisualSynth}}.

\begin{table}
\centering
\setlength\abovecaptionskip{-10pt}
\setlength{\tabcolsep}{3pt}
\renewcommand{\arraystretch}{1.1}
\begin{tabular}{lcccc}
\toprule
\textbf{Class} & \multicolumn{2}{c}{LE$^\circ$ $\downarrow$} & \multicolumn{2}{c}{LR $\uparrow$} \\
\cmidrule(lr){2-3} \cmidrule(lr){4-5}
& Before & Now & Before & Now \\
\midrule
Speech (F) & 24.80 & \cellcolor{lightred}26.87 & 0.75 & 0.75 \\
Speech (M) & 19.64 & \cellcolor{darkblue}15.58 & 0.68 & \cellcolor{lightred}0.66 \\
Clapping & 16.17 & \cellcolor{mediumred}19.03 & 0.47 & \cellcolor{darkblue}0.71 \\
Telephone & 24.31 & \cellcolor{mediumblue}20.03 & 0.60 & \cellcolor{lightblue}0.65 \\
Laughter & 19.06 & \cellcolor{darkblue}15.41 & 0.35 & \cellcolor{darkred}0.25 \\
Appliance & 21.59 & \cellcolor{lightblue}20.68 & 0.74 & \cellcolor{mediumred}0.66 \\
Footsteps & 18.57 & \cellcolor{darkred}30.63 & 0.42 & \cellcolor{darkred}0.14 \\
Door & 10.05 & \cellcolor{lightred}12.12 & 0.14 & \cellcolor{mediumblue}0.26 \\
Music & 32.14 & \cellcolor{lightblue}31.92 & 0.68 & \cellcolor{mediumred}0.58 \\
Instrument & 14.78 & \cellcolor{mediumred}19.08 & 0.60 & \cellcolor{lightred}0.59 \\
Water tap & 23.48 & \cellcolor{lightred}25.18 & 0.04 & \cellcolor{darkblue}0.62 \\
Bell & 23.93 & \cellcolor{darkred}33.95 & 0.45 & \cellcolor{darkblue}0.69 \\
Knock & 15.63 & \cellcolor{lightblue}14.62 & 0.07 & \cellcolor{darkblue}0.79 \\
\bottomrule
\end{tabular}
\setlength\abovecaptionskip{-20pt}
\caption{Per-class performance comparison of AV SELDnet-YOLOv8 \cite{roman2024enhanced} (Before) and AV SELDnet-YOLOv8 trained with SELDVisualSynth (Now) across the 13 STARSS23 sound event classes. Color intensity reflects degree of change (blue denotes improvement and red denotes deterioration). Note the gains in localization recall for classes like `Door', `Water tap', `Bell' and `Knock'.}
\label{tab:performance_metrics}
\end{table}

\section{Methods}

\subsection{Dataset}

Our training dataset includes synthetic audio-visual data that we generate using the following pipeline:

\textbf{Audio synthesis:} We use SpatialScaper \cite{roman2024spatial} to generate 2,000 First Order Ambisonics (FOA) audio clips, each lasting 60 seconds, sampled at 24 kHz, and simulated in 14 different rooms \cite{politis2021dataset, orhun_olgun_2019_2635758, mckenzie2021dataset, gotz2021dataset, chesworth2024room, schneiderwind2019data}. Consistent with the STARSS dataset \cite{politis2022starss22,shimada2023starss23}, SpatialScaper generates metadata including spatiotemporal DoA information and class annotations for 13 target classes. The audio clips are generated with a maximum polyphony of three simultaneous sound events.

\textbf{Video synthesis:} For the visual modality, we propose SELDVisualSynth, a tool to create 360$^\circ$ synthetic videos based on the metadata files generated by SpatialScaper. SELDVisualSynth includes a collection of video and image assets categorized according to the 13 classes in STARSS, spatialized on top of newly-collected 360$^\circ$ background images. The video and image assets are resized to 50×50 pixel squares (i.e. tiles) and background images at 1920×960 resolution.

\begin{table*}[h]
 \begin{center}
 \begin{tabular}{lccccccc}
   \toprule
    \multicolumn{4}{c}{\textbf{Model Configuration}} & \multicolumn{4}{c}{\textbf{Performance Metrics}} \\
    \cmidrule(lr){1-4} \cmidrule(lr){5-8}
    Model Type & Visual Detector & Input Features & Data Augmentation & ER$_{20^\circ}$ $\downarrow$ & F$_{20^\circ}$ $\uparrow$ & LE$^\circ$ $\downarrow$ & LR $\uparrow$ \\ 
   \midrule
    AO SELDnet \cite{adavanne2018sound} & - & FOA & ACS & \textbf{0.57} & 29.9 & \underline{21.6} & \underline{47.7} \\ 
    AV SELDnet \cite{shimada2023starss23} & YOLOX & FOA + Video & - & 1.07 & 14.3 & 48.0 & 35.5 \\ 
    AV SELDnet \cite{shimada2023starss23} & YOLOX & FOA + Video & ACS + VPR & 1.37 & 15.0 & 40.62 & 40.0 \\ 
    AV SELDnet \cite{roman2024enhanced} & YOLOv8 & FOA + Video & ACS + VPR & 0.63 & \underline{30.9} & \textbf{20.3} & 46.1 \\ 
   \midrule
    \textbf{AV SELDnet} & YOLOv8 & FOA + Video & \textbf{SELDVisualSynth Data} & \underline{0.62} & \textbf{33.2} & 21.9 & \textbf{56.4} \\ 
   \bottomrule
   \end{tabular}
\end{center}
\setlength\abovecaptionskip{-10pt}
 \caption{AO and AV SELDnet performance on the `test' split from the STARSS23 development set. The Data Augmentation column indicates whether training data was augmented using audio channel swapping (ACS), video pixel rotation (VPR), or by including the data synthesized by our SELDVisualSynth pipeline. Bold and underlined numbers indicate best and second best score for each metric.}
 \label{tab:starss_results}
\end{table*}

During video generation, for each sound event in the SpatialScaper metadata, SELDVisualSynth randomly selects a corresponding visual representation in the form of a tile that matches the sound event class. Each tile is then positioned in the appropriate time and pixel coordinates within the video background, according to the SELD DoA labels. The resulting video is synchronized with the audio stream. 
We generate a total of 2,000 video clips, each corresponding to an FOA audio clip that we generated with SpatialScaper.

\subsection{Models trained and evaluated}

For the audio-only modality, we use the SELDnet \cite{shimada2023starss23, adavanne2018sound} baseline model from the DCASE Challenge, Task 3. SELDnet is equipped with multi-ACCDOA \cite{shimada2022multi}, allowing it to simultaneously infer the presence, class, and spatial coordinates of up to three sound events. The model also includes two multi-head self-attention (MHSA) layers \cite{sudarsanam2021assessment} to enhance its ability to capture temporal dependencies.
For the audio-visual modality, we use the audio-visual version of SELDnet \cite{shimada2023starss23}. This architecture consists of an audio and a vision branch: the audio branch is identical to that of the audio-only baseline. The vision branch utilizes YOLOX \cite{ge2021yolox} as an object detection feature extractor, primarily detecting `human' objects within bounding boxes. Audio-visual models in our benchmark were also trained with data that was augmented using audio channel swapping (ACS) and video pixel rotation (VPR) techniques \cite{wang2023four}, but our proposed approach did not have to use these data augmentation techniques to attain good performance.

\subsection{Metrics}

We employ the SELD metrics proposed by the DCASE Challenge. Two metrics relate to DoA estimation: F1-score (F$_{20^\circ}$) and error rate (ER$_{20^\circ}$). F$_{20^\circ}$ is calculated from location-aware precision and recall. ER$_{20^\circ}$ is the sum of insertion, deletion, and substitution errors divided by the total number of inferred audio frames. The other two metrics relate to class-aware localization: localization error (LE) in degrees and localization recall (LR). LE is the average angular difference between each class prediction and its label. LR is the true positive rate of instantaneous detections out of the total annotated sounds.

\subsection{Training procedure}

We use the audio-visual SELDnet architecture with YOLOv8 \cite{roman2024enhanced}. The main difference lies in our training data: we use the development set from STARSS23 and we add 2,000 audio and video clips generated using our procedure. Similar to \cite{roman2024enhanced}, we validate using the `test' split from STARSS23.

\section{Results}

Tables \ref{tab:performance_metrics} and \ref{tab:starss_results} present results obtained on the STARSS23 `test' split. We compare audio-only (AO) and audio-visual (AV) SELDnet baselines to our AV SELDnet with YOLOv8, trained on SpatialScaper synthetic audio and SELDVisualSynth videos.

Table \ref{tab:starss_results} shows that the AO SELDnet outperforms all AV SELDnet implementations in the error rate metric, followed closely by our proposed approach. Without any data augmentation, AV SELDnet achieves an ER$_{20^\circ}$ of 1.07 (higher is worse) and a low F$_{20^\circ}$ of 14.3. Adding ACS and VPR augmentations failed to improve performance, increasing the ER$_{20^\circ}$ to 1.37.

The AV SELDnet model achieves significant improvements with the YOLOv8 detector with ACS and VPR data augmentation \cite{roman2024enhanced}. This configuration achieves competitive results with an ER$_{20^\circ}$ of 0.63 and an F$_{20^\circ}$ of 30.9, approaching the performance of the audio-only baseline, and even reducing localization error to 20.3 degrees.

Our proposed AV SELDnet trained with SELDVisualSynth further improves ER$_{20^\circ}$ to 0.62 and attains the highest F$_{20^\circ}$ of 33.2. Most notably, it demonstrates the best LR of 56.4, an 18.2\% improvement over the audio-only baseline and a 22.3\% improvement over the best prior AV SELDnet implementation. The localization error remains competitive at 21.9 degrees. 

Table \ref{tab:performance_metrics} presents a class-wise performance comparison (also on the STARSS23 dataset) of the two AV SELDnet models in the bottom two rows of Table \ref{tab:starss_results}. Overall, the LE remains similar across both models, except for `Footsteps' and `Bell'. We hypothesize that this was perhaps due to the challenge of sourcing image and video assets for these classes.

These results demonstrate that the diverse audio visual data, produced by our 
SELDVisualSynth data generation approach, can significantly boost SELD system performance. Our proposed system even outperforms or is competitive against the other models in the benchmark without needing the ACS and VPR augmentations. 
This suggests that the synthetic data that we produce allows models to leverage complementary audio-visual information, to enhance the system's ability to detect, localize and classify sound events.

\section{Conclusion}

We introduced SELDVisualSynth, a synthetic data generator tool for audio-visual SELD. We incorporate naturalistic background images, on top of which video and image tiles of sounding objects are positioned with and precise temporal alignment with audio. This enables video synthesis to train multimodal SELD models and improve their performance. Experimental results demonstrate that models trained with SELDVisualSynth achieve superior localization recall and competitive localization error without having to rely on other data augmentation techniques. These findings highlight the potential of synthetic audio-visual approaches to advance SELD research and provide a robust foundation for training SELD systems.

\pagebreak
\section{Acknowledgments}
This research was partially supported by  Mexico's National Scholarship for Graduate Studies from the Secretariat of Science, Humanities, Technology, and Innovation (SECIHTI).
\bibliographystyle{IEEEtran}
\bibliography{mybib}

\end{document}